\documentclass[11pt]{article}
\usepackage{amssymb}
\usepackage{amsmath}
\usepackage{amscd}
\usepackage{latexsym}

\catcode`@=11
\renewcommand{\section}{\@startsection{section}{1}{0pt}{\medskipamount}
{\medskipamount}{\large\bf}}
\numberwithin{equation}{section}
\catcode`@=12
\def\a{\alpha}
\def\b{\beta}

\def\de{\delta}

\def\ve{\varepsilon}

\def\la{\lambda}
\def\m{\mu}
\def\n{\nu}

\def\s{\sigma}

\def\1{\dot 1}
\def\2{\dot 2}

\newcommand{\C}{\mathbb C}
\newcommand{\R}{\mathbb R}

\newcommand{\Acal}{{\cal A}}

\newcommand{\Gcal}{{\cal G}}
\newcommand{\Ncal}{{\cal N}}

\newcommand{\Fcal}{{\cal F}}

\def\im{\textrm{i}}

\def\N2{$N{=}2$}

\def\diff{\textrm{d}}
\def\tr{\textrm{tr}}
\def\sfrac#1#2{{\textstyle\frac{#1}{#2}}}
\def\>{\rangle}
\def\<{\langle}
\def\+{\dagger}
\def\={\ =\ }
\def\und{\qquad\textrm{and}\qquad}

\begin{document}

\begin{titlepage}
\setcounter{page}{0}
\begin{flushright}
ITP--UH--14/08\\
\end{flushright}

\vskip 2.0cm

\begin{center}

{\Large\bf  Yang-Mills Instantons and Dyons on Group Manifolds
}

\vspace{12mm}

{\large Tatiana A. Ivanova${}^*$  \ and \ Olaf Lechtenfeld${}^\+$
}
\\[8mm]
\noindent ${}^*${\em Bogoliubov Laboratory of Theoretical Physics, JINR\\
141980 Dubna, Moscow Region, Russia}\\
{Email: ita@theor.jinr.ru}
\\[8mm]
\noindent ${}^\+${\em Institut f\"ur Theoretische Physik,
Leibniz Universit\"at Hannover \\
Appelstra\ss{}e 2, 30167 Hannover, Germany }\\
{Email: lechtenf@itp.uni-hannover.de}

\vspace{12mm}

\begin{abstract}
\noindent
We consider Euclidean SU($N$) Yang-Mills theory on the space $G{\times}\R$,
where $G$ is a compact semisimple Lie group, and introduce first-order
BPS-type equations which imply the full Yang-Mills equations.
For gauge fields invariant under the adjoint $G$-action these BPS~equations
reduce to first-order matrix equations, to which we give instanton solutions.
In the case of $G={}$SU(2)${}\cong S^3$, our matrix equations are recast
as Nahm equations, and a further algebraic reduction to the
Toda chain equations is presented and solved for the SU(3) example.
Finally, we change the metric on $G{\times}\R$ to Minkowski and construct
finite-energy dyon-type Yang-Mills solutions. The special case of
$G={}$SU(2)${\times}$SU(2) may be used in heterotic flux compactifications.
\end{abstract}

\end{center}
\end{titlepage}

\section{Introduction and summary}

\noindent
Instantons~\cite{1} play an important role in modern gauge field 
theories~\cite{2,3}.
They are nonperturbative BPS configurations in four Euclidean dimensions 
solving the first-order self-duality equations and forming a subset 
of solutions to the full Yang-Mills equations. 
On $\R^4$ the construction of instantons was described systematically in the 
framework of the twistor approach~\cite{4} and by the ADHM method~\cite{5}. 
They were also generalized to noncommutative $\R^4$ (see e.g.~\cite{6}
and references therein) which resolved the problem of `zero size' instantons.

On the other hand, in the celebrated AdS/CFT correspondence~\cite{7} 
$\Ncal{=}4$ supersymmetric Yang-Mills theory appears on $S^3{\times}\R$,
which is the boundary of AdS$_5$. 
Hence, it is of interest to study instanton configurations on $S^3{\times}\R$
with Euclidean signature and on its further compactification $S^3{\times}S^1$
(thermal time circle)~\cite{8}. For the group SU(2) embedded into SU($N$), 
their construction was discussed e.g.\ in~\cite{9,10}.
Furthermore, for $N{>}2$, \cite{10} reduced the self-duality equations 
on $S^3{\times}\R$ to Toda-like equations and gave some explicit solutions.

In space-time dimensions higher than four, BPS~configurations can still 
be found as solutions to first-order equations, known as generalized 
self-duality equations~\cite{11}-\cite{12b}.
These appear in superstring compactifications as conditions for the survival 
of at least one supersymmetry~\cite{13}. 
Various solutions to these first-order equations were found
e.g.\ in~\cite{11,12b,14,15}, and their noncommutative generalizations 
have been considered e.g.\ in~\cite{16,17}, 
alongside with their brane interpretation.

In this paper, we revisit the first-order Yang-Mills BPS-type equations 
on the Euclidean group manifold $G{\times}\R$, where $G$ is a compact 
semisimple Lie group. For Ad${}_G$-invariant gauge potentials this system 
reduces to a particular matrix mechanics admitting finite-action (instanton) 
solutions. 
For the case of $G={}$SU(2)${}\cong S^3$, which was also studied in~\cite{10},
we reformulate the matrix equations as Nahm equations and reduce these further
to the Toda chain equations, whose explicit solutions we present for the 
SU(3)~example.
Generically, the action for the Toda chain solutions turns out to be infinite,
hence these configurations describe an instanton gas or liquid (cf.~\cite{17'}).

Considering Minkowski signature on the space $G{\times}\R$, we also construct 
non-BPS dyon solutions of the Yang-Mills equations on this space, 
generalizing those on $S^3{\times}\R$ found in~\cite{18}.
Their energy is proportional to the volume of the group~$G$, 
hence is finite for compact Lie groups. 

As an application of our results one may consider the special case of 
$G={}$SU(2)${\times}$SU(2), which appears in the compactification 
of heterotic strings on Ad$S_4{\times}S^3{\times}S^3$ with an SU(3) structure 
(nonvanishing $H$-field) on the internal six-manifold
(see e.g.~\cite{18'} and references therein). 
It would be interesting to generalize our solutions to such a general setting.

\bigskip

\section{Instantons in Yang-Mills theory on $G\times\R$}

\noindent
Let $\Gcal$ be a Lie algebra of a compact semisimple Lie group $G$. 
We assume that the structure constants $f_{abc}$ of $G$ are normalized 
so that $f_{acd}f_{bcd}=2\de_{ab}$ for $a,b,\ldots{=}1,\ldots,n{-}1$ and 
$n{=}\dim\Gcal{+}1$. Consider the vector space $\R^n=\Gcal\oplus\R$ 
with the Euclidean metric $(\de_{\m\n})=(\de_{ab},\de_{nn})$
with $\m,\n,\ldots{=}1,\ldots,n$. It is obvious that $\Gcal\oplus\R$ 
can be regarded as a Lie algebra with the commutation relations
\begin{equation}
[I_a, I_b]=f_{abc}\, I_c \und [I_a, I_n]=0\ ,
\end{equation}
where the $I_a$ are generators of $G$ and $I_n$ generates the translations
along~$\R$.
So with a coordinate $\tau$ on $\R$ we multiply
$(g_1,\tau_1)\cdot (g_2,\tau_2)=(g_1g_2,\,\tau_1{+}\tau_2)$ on $G\times \R$.

On $G\times\R$ we consider left-invariant one-forms $\{e^a,e^n\}$ 
defined by the equations
\begin{equation}\label{ea}
g^{-1}\diff g \= -\sfrac{1}{R}\, e^a\, I_a \und  -\diff\tau \= e^n\ ,
\end{equation}
where $g\in G$ and $R$ is a dimensional parameter characterizing 
the `size' of $G$. These one-forms satisfy the Maurer-Cartan equations
\begin{equation}
\diff e^a -\sfrac{1}{2R}\, f^a_{\ \,bc}\, e^b\wedge e^c\=0 \und \diff e^n \= 0
\end{equation}
which easily follow from the fact that 
$g^{-1}\diff g$ is the canonical flat connection on $G$. 

Let us now specialize to Ad${}_G$-invariant $su(N)$-valued gauge potentials
$\Acal$, which in the `temporal gauge' $\Acal_\tau=0$ take the form
\begin{equation}\label{Acal}
\Acal(g,\tau) \= X_a(\tau)\,e^a(g) 
\qquad\textrm{with}\quad X_a(\tau)\in su(N)\ .
\end{equation}
The corresponding gauge field reads
\begin{equation}\label{Fcal}
\Fcal\ \equiv\ \diff\Acal\ +\ \Acal\wedge\Acal \= \dot X_a\,e^a\wedge e^n\ +\
\sfrac12\bigl(\sfrac{1}{R}\,f_{abc}\,X_c+[X_a,X_b]\bigr)\,e^a\wedge e^b\ ,
\end{equation}
where $\dot X_a:=\diff X_a/\diff\tau$. 
{}From (\ref{Fcal}) we extract the components $\Fcal_{\m\n}$ of $\Fcal$,
\begin{equation}\label{Fab}
\Fcal_{ab}\=\sfrac{1}{R}\, f_{abc}\,X_c\ +\ [X_a, X_b] 
\und \Fcal_{cn}\=\dot X_c\ ,
\end{equation}
in the nonholonomic basis $\{e^\m\}=\{e^a, e^n\}$ of one-forms on $G\times\R$.
Note that the metric on the group manifold $G\times\R$ has the form
\begin{equation}\label{metric}
\diff s^2 \= \de_{\m\n}\, e^{\m}e^{\n} \=\de_{ab}\, e^ae^b + (e^n)^2 \ =:\
R^2\,\diff\Omega^2_{n-1} +\,\diff\tau^2\ .
\end{equation}

On $\R^n=\Gcal\oplus\R$ we introduce the following completely antisymmetric 
four-index tensor $T_{\m\n\la\s}$:
\begin{equation}\label{T}
T_{abcd}\=0 \und T_{abcn}\=f_{abc}\ .
\end{equation}
Consider now the first order (BPS) equations
\begin{equation}\label{bps}
\sfrac12\, T_{\m\n\la\s}\, \Fcal_{\la\s}= \Fcal_{\m\n}\ ,
\qquad\textrm{i.e.}\quad \Fcal_{ab}\= f_{abc}\,\Fcal_{cn}\ ,
\end{equation}
written in the nonholonomic basis $\{e^{\m}\}$ on $G{\times}\R$ 
defined by (\ref{ea}). 
These BPS equations generalize the self-duality equations~\cite{1} on $\R^4$. 
The group~$G$ can be embedded into the rotation group SO$(n{-}1)$
and acts on $\Gcal$ via the adjoint representation Ad${}_G$, leaving invariant
the tensor $T_{\m\n\la\s}$ and therefore the equations~(\ref{bps}).
{}From (\ref{Fcal}) it follows that (\ref{bps}) is equivalent to 
\begin{equation}\label{dotX}
f_{abc}\,\dot X_c \=\sfrac{1}{R}\,f_{abc}\,X_c\ +\ [X_a, X_b]\ .
%\qquad\Rightarrow\qquad
%\dot X_a \=\sfrac{1}{R}\, X_a\ +\  \sfrac{1}{2}\, f_{abc}[X_b, X_c]\ .
\end{equation}
Note that the full Yang-Mills equations on $G\times\R$ after substitution
of (\ref{Acal})-(\ref{Fab}) reduce to the matrix equations
\begin{equation}\label{ddotX}
\ddot X_a \=\sfrac{1}{R^2}\, X_a\ +\ \sfrac{3}{2R}\, f_{abc}\,[X_b, X_c]\ +\
\sfrac{1}{2}\, f_{abc}f_{cde}\,\bigl[X_b, [X_d, X_e]\bigr]\ ,
\end{equation}
and each solution of (\ref{dotX}) satisfies (\ref{ddotX}) as can be seen after
differentiation (\ref{dotX}) and multiplication by the structure constants.

To find a solution of (\ref{dotX}) we embed $\Gcal$ into $su(N)$ 
with the ansatz (cf.~\cite{10, 18})
\begin{equation}\label{Xa}
X_a(\tau)\=-\bigl(\phi(\tau) + \sfrac{1}{2R}\bigr)\, I_a\ ,
\end{equation}
where $\phi(\tau)\in\R$ and the $I_a$ generate an $N$-dimensional 
representation of~$G$.
Substituting (\ref{Xa}) into (\ref{dotX}), we arrive at
\begin{equation}\label{dotphi}
\dot\phi \= \sfrac{1}{4R^2} - \phi^2
\end{equation}
whose solution is the kink
\begin{equation}\label{phi}
\phi \= \sfrac{1}{2R}\, \tanh\bigl(\sfrac{\tau}{2R}\bigr)\ ,
\end{equation}
where one can shift $\tau\mapsto\tau{-}\tau_0$ due to translational invariance.

Inserting (\ref{phi}) into (\ref{Acal}) and (\ref{Fcal}), we obtain 
the corresponding solution of the Yang-Mills equations on $G\times\R$,
\begin{equation}\label{2.16}
\Acal \= -\sfrac{1}{2R}\bigl(\tanh(\sfrac{\tau}{2R})+1\bigr)\, e^aI_a
\und
\Fcal \= \sfrac{1}{4R^2\cosh^2(\frac{\tau}{2R})} 
\bigr(e^a\wedge e^n + \sfrac{1}{2} f^a_{\ bc}\,e^b\wedge e^c\bigl)\, I_a\ ,
\end{equation}
which is an instanton since
\begin{equation}\label{2.17}
\Acal\to 0 \quad\textrm{for}\quad \tau\to -\infty
\und
\Acal\to -\sfrac{1}{R}\, e^aI_a=g^{-1}\diff g 
\quad\textrm{for}\quad \tau\to +\infty\ ,
\end{equation}
where $g:\ G\to\ $SU($N$) is a degree one map.
Furthermore, the action functional for this solution is
\begin{equation}\label{2.18}
S\=-\int_{G\times\R}\tr (\Fcal\wedge *\Fcal ) \=
2\,c\,(n{-}1)\,\textrm{Vol}(G)\int^\infty_{-\infty}\!\!\diff\tau\ \dot\phi^2 \=
\sfrac13\,c\,(n{-}1)\,\textrm{Vol}(G)\,R^{-3}\ <\ \infty\ ,
\end{equation}
where $c$ is the Dynkin index of the $N$-dimensional $G$-representation
generated by~$I_a$.

\bigskip

\section{Yang-Mills theory on $S^3\times\R$ and Toda chain equations}

\noindent
In the special case of $\Acal\in su(2)$ on $G={}$SU(2) with size~$R$, 
i.e.~instantons on $S^3_{2R}{\times}\R$, from (\ref{2.18}) with 
$c{=}\sfrac12$, $n{-}1{=}3$ and Vol$(G){=}2\pi^2(2R)^3$ we obtain $S{=}8\pi^2$, 
which is exactly the action of the one-instanton BPST~solution.
For this case, the matrix equations (\ref{dotX}) read
\begin{equation}\label{3.1}
\dot X_a\=\sfrac{1}{R}\, X_a\ +\ \sfrac{1}{2}\ve_{abc}\,[X_b, X_c]
\qquad\textrm{with}\quad a,b,\ldots = 1,2,3\ .
\end{equation}
To further simplify we introduce a new variable $r\in (0,\infty)$,
\begin{equation}\label{3.2}
r\=R\, \exp\bigl(\sfrac{\tau}{R}\bigr)
\qquad\Leftrightarrow\qquad
\tau \= R\,\log\bigl(\sfrac{r}{R}\bigr)
\end{equation}
so that
\begin{equation}\label{3.3}
\diff s^2\=\de_{ab}e^a e^b+\diff\tau^2 \=\diff\tau^2 +R^2\,\diff\Omega^2_3 \=
\sfrac{R^2}{r^2}\bigl(\diff r^2+r^2\,\diff\Omega^2_3\bigr)\ .
\end{equation}
At the same time, we redefine our matrix functions (cf.~\cite{19}) as
\begin{equation}
X_a(\tau)\ \mapsto\ Y_a(r)\ :=\ 
\exp\bigl(-\sfrac{\tau}{R}\bigr )\,X_a\bigl(\tau(r)\bigr)\ ,
\end{equation}
which transforms (\ref{3.1}) to the well-known Nahm equations~\cite{20}
\begin{equation}\label{3.5}
\sfrac{\diff}{\diff r} Y_a \= \sfrac{1}{2}\ve_{abc}\,[Y_b\,,Y_c]\ .
\end{equation}
 
Nahm's equations admit a Lax representation, i.e.~they can be written as
\begin{equation}\label{3.6}
\sfrac{\diff }{\diff r}L(\zeta) \= \bigl[L(\zeta )\,, M(\zeta )\bigr]\ ,
\end{equation}
with
\begin{equation}
L(\zeta )\= (1{+}\zeta^2)\,Y_1 + \im\,(1{-}\zeta^2)\,Y_2 - 2\im\,\zeta\,Y_3
\und M(\zeta )\= \zeta (Y_1{-}\im\,Y_2) - \im\,Y_3 \ ,
\end{equation}
where $\zeta\in\C P^1$ is an extra `spectral' parameter. Therefore,
to (\ref{3.5}) one can apply the integrable systems' machinery for constructing
solutions, conserved charges (see e.g.~\cite{19}) etc.. In fact, a general
solution of~(\ref{3.5}) can be given in terms of theta functions.

Here, we are interested in a special class of solutions which arise 
by reducing~(\ref{3.6}) to (periodic) Toda lattice equations 
(see e.g.~\cite{12b, 21}) and to Toda chain equations, in particular.
Namely, let us consider the Chevalley basis $\{H_\a , E_\a , E_{-\a}\}$
for the Lie algebra $su(N)$,
\begin{equation}\label{3.8}
[H_\a , H_\b ]=0\, ,\quad 
[E_\a , E_{-\b}]=\de_{\a\b}\,H_\b\, ,\quad 
[H_\a , E_\b ]=K_{\a\b}E_\b\, ,\quad
[H_\a , E_{-\b} ]=-K_{\a\b}\,E_{-\b}\ ,
\end{equation}
where $(K_{\a\b})$ is the Cartan matrix and $\a,\b,\ldots=1,\ldots,N{-}1$.
We then put
\begin{equation}\label{3.9}
Y_1\=\frac12\sum\limits_{\a=1}^{N-1} f_\a(r)\,(E_\a{-}E_{-\a})\ ,\quad
Y_2\=\frac{\im}{2}\sum\limits_{\a=1}^{N-1} f_\a(r)\,(E_\a{+}E_{-\a})\ ,\quad
Y_3\=\frac{\im}{2}\sum\limits_{\a=1}^{N-1} h_\a(r)\, H_\a \ ,
\end{equation}
introducing real-valued functions $f_\a$ and $h_\a$.
In addition, we relate the latter via `potentials' $\phi_\a$ as
\begin{equation}\label{3.10}
f_\a \= \exp(\sfrac12\phi_\a) \und
h_\a \= \sum_{\b=1}^{N-1} K_{\a\b}^{-1}\,\sfrac{\diff}{\diff r}\phi_\b\ ,
\end{equation}
where $K^{-1}$ is the inverse Cartan matrix.
With this, the Nahm equations (\ref{3.5}) as well as 
their Lax representation (\ref{3.6}) turn into the Toda chain equations
\begin{equation}\label{3.11}
\sfrac{\diff^2}{\diff r^2}\phi_\a \=\sum_{\b=1}^{N-1}K_{\a\b}\,\exp(\phi_\b)\ .
\end{equation}

The general solution to~(\ref{3.11}) is known (see e.g.~\cite{22}).
As an example we display it for $N{=}3$:
\begin{equation}\label{3.12}
\begin{aligned}
f_1&\= A_1\exp(A_1r{+}B_1)\frac{\sqrt\Phi}{\Psi}\ , \qquad
f_2 \= A_2\exp(A_2r{+}B_2)\frac{\sqrt\Psi}{\Phi}\ ,\\
h_1&\= \frac{4A_1{+}2A_2}{3}\ +\ \frac{A_1\exp(2A_1r{+}2B_1)}{\Psi}
\Bigl (1-\frac{A_1}{A_1{+}A_2}\exp(2A_2r + 2B_2)\Bigr )\ ,\\
h_2&\= \frac{2A_1{+}4A_2}{3}\ +\ \frac{A_2\exp(2A_2r{+}2B_2)}{\Phi}
\Bigl (1-\frac{A_2}{A_1{+}A_2}\exp(2A_1r{+}2B_1)\Bigr )\ ,\\
\Psi &\= 1\ -\ \exp(2A_1r{+}2B_1)\ +\ 
\frac{A_1^2}{(A_1{+}A_2)^2}\exp(2A_1r{+}2B_1) \exp(2A_2r{+}2B_2)\ ,\\
\Phi &\= 1\ -\ \exp(2A_2r{+}2B_2)\ +\ 
\frac{A_2^2}{(A_1{+}A_2)^2}\exp(2A_1r{+}2B_1) \exp(2A_2r{+}2B_2)\ ,
\end{aligned}
\end{equation}
where $A_1, A_2, B_1, B_2$ are arbitrary constants.
Substituting (\ref{3.2}), (\ref{3.10}) and (\ref{3.12}) into (\ref{Acal}) 
and~(\ref{Fcal}), we obtain a smooth solution of the self-dual Yang-Mills 
equations in four dimensions. However, the field strength~$\Fcal$ does not 
vanishing for $\tau$ or $r\to\infty$. Therefore,
the action and topological charge for such solutions are infinite, 
i.e.~they describe an instanton gas or instanton liquid (see e.g.~\cite{17'}).
Only for limiting values of the parameters $A_1, A_2, B_1, B_2$, 
the action functional for the solution~(\ref{3.12}) becomes finite -- 
in this case it reduces to the one-instanton solution discussed earlier.

\bigskip

\section{Dyons in Yang-Mills theory on $G\times\R$}

\noindent
Finally let us change the signature of $G\times\R$ from Euclidean to 
Minkowski by choosing on $\R$ a coordinate
\begin{equation}
x^0=t=-\im\tau \qquad\textrm{so that}\quad e^0=\,\diff t\ .
\end{equation}
For $\Acal$ we copy the ansatz (\ref{Acal}) but now
\begin{equation}\label{4.1}
\Fcal \=\sfrac{\diff}{\diff t}\,X_a\, e^0\wedge e^a\ +\ \sfrac{1}{2}
\bigl(\sfrac{1}{R}\,f_{abc}\,X_c + [X_a, X_b]\bigr)\, e^a\wedge e^b\ .
\end{equation}
After substituting (\ref{Acal}) and (\ref{4.1}) into the Yang-Mills
equations on $G\times\R$, we obtain the matrix equations
\begin{equation}\label{4.2}
\sfrac{\diff^2}{\diff t^2}\,X_a\ +\ \sfrac{1}{R^2}\,X_a\ +\ 
\sfrac{3}{2R^2}\,f_{abc}\,[X_b, X_c]\ +\ 
\sfrac{1}{2}\,f_{abc}f_{cde}\,\bigl[X_b, [X_d, X_e]\bigr] \=0\ ,
\end{equation}
which also follow from (\ref{ddotX}) by $\tau\mapsto\im t$.
Note that for the Minkowski signature we cannot write a first-order form
for these equations inside~$su(N)$.

Still, the previous ansatz (\ref{Xa}) reduces the equations~(\ref{4.2}) to
\begin{equation}\label{4.3}
\sfrac{\diff^2}{\diff t^2}\,\phi\ -\ \sfrac{1}{2R^2}\,\phi + 2\phi^3 \=0\ ,
\end{equation}
which is solved by
\begin{equation}\label{4.4}
\phi \=\bigl[ \sqrt{2}\,R\,\cosh(\sfrac{t}{\sqrt{2}\,R}))\bigr]^{-1}\ .
\end{equation}
For the case of $G={}$SU(2), such solutions were discussed in~\cite{18}.

By inserting (\ref{4.4}) into (\ref{Acal}) and (\ref{4.1}), 
we arrive at a dyon configuration
\begin{subequations}\label{4.5}
\begin{eqnarray}
\Acal &=& -\frac{1}{2R}
\Bigl(1+\frac{\sqrt{2}}{\cosh(\frac{t}{\sqrt{2}\,R})}\Bigr)\,e^aI_a\ ,\\
\Fcal &=& \biggl(\frac{1}{2R^2}
\frac{\sinh(\frac{t}{\sqrt{2}R})}{\cosh^2(\frac{t}{\sqrt{2}R})}\,e^0\wedge e^a\
+\ \frac{1}{8R^2}\,\Bigl(\frac{2}{\cosh^2(\frac{t}{\sqrt{2}\,R})}-1\Bigr)\, 
f_{abc}\,e^b\wedge e^c\biggr)\,I_a\ ,
\end{eqnarray}
\end{subequations}
{}from which we extract the components
\begin{equation}\label{4.6}
\Fcal_{0a}\= \frac{1}{2R^2}\,
\frac{\sinh(\frac{t}{\sqrt{2}R})}{\cosh^2(\frac{t}{\sqrt{2}R})}\,I_a
\und
\Fcal_{ab}\= \frac{1}{4R^2}\,
\frac{2-\cosh^2(\frac{t}{\sqrt{2}\, R})}{\cosh^2(\frac{t}{\sqrt{2}\, R})}\, 
f_{abc}\,I_c\ .
\end{equation}
Hence, the energy density becomes
\begin{equation}\label{4.7}
{\cal E} \= -\tr\,(2\Fcal_{0a}\Fcal_{0a}+\Fcal_{ab}\Fcal_{ab}) \=
\sfrac18\,c\,(n{-}1)\,R^{-4}\ ,
\end{equation}
where $c$ is the Dynkin index of the embedding representation.
Integrating over the compact Lie group~$G$ we finally obtain the energy
\begin{equation}\label{4.8}
E\=\sfrac18\,c\,(n{-}1)\,\textrm{Vol}(G)\,R^{-4} \ <\ \infty\ .
\end{equation}

\bigskip

\noindent
{\bf Acknowledgements}

\medskip

\noindent
The authors are grateful to A.D.~Popov for fruitful discussions and useful
comments.
T.A.I.~acknowledges the Heisenberg-Landau program and the Russian Foundation
for Basic Research (grant 06-01-00627-a) for partial support and the Institut
f\"ur Theoretische Physik der Leibniz Universit\"at Hannover for its
hospitality. The work of O.L.~is partially supported by the Deutsche
Forschungsgemeinschaft.
\bigskip
%\newpage

\end{document}